\documentclass[twocolumn,showpacs,preprintnumbers,amsmath,amssymb,prb,aps]{revtex4}

\usepackage{epsfig}% Include figure files
\usepackage {graphicx}% Include figure files
\usepackage{dcolumn}% Align table columns on decimal point
\usepackage{bm}% bold math\part{title}
\usepackage{tabularx}
\begin{document}
\title{Electronic structure studies on single crystalline Nd$_2$PdSi$_3$, an exotic Nd-based intermetallic: Evidence for Nd 4$f$ hybridization}
\author{Kalobaran Maiti$^1$}
\altaffiliation{Corresponding author: kbmaiti@tifr.res.in}
\author{Tathamay Basu$^1$}
\altaffiliation{Present address: Department of Physics and Astronomy, Michigan State University, East Lansing, Michigan 48824, USA}
\author{Sangeeta Thakur$^1$}
\author{Nishaina Sahadev$^1$}
\author{Deepnarayan Biswas$^1$}
\author{Ganesh Adhikary$^1$}
\author{C. G. F. Blum$^2$}
\author{W. L\"{o}ser$^2$}
\author{E. V. Sampathkumaran$^1$}

\affiliation{$^1$ Department of Condensed Matter Physics and Materials Science, Tata Institute of Fundamental Research, Homi Bhabha Road, Colaba, Mumbai - 400 005, India}
\affiliation{$^2$ IFW Dresden, Leibniz-Institut f\'{u}r Festk\"{o}rper- und Werkstoffforschung, Helmholtzstr. 20, D-01171, Dresden, Germany}

\date{\today}

\begin{abstract}
In the series R$_2$PdSi$_3$, Nd$_2$PdSi$_3$ is an anomalous compound in the sense that it exhibits ferromagnetic order unlike other members in this family. The magnetic ordering temperature is also unusually high compared to the expected value for a Nd-based system, assuming 4$f$ localization. Here, we have studied the electronic structure of single crystalline Nd$_2$PdSi$_3$ employing high resolution photoemission spectroscopy and ab initio band structure calculations. Theoretical results obtained for the effective electron correlation strength of 6 eV corroborate well with the experimental valence band spectra. While there is significant Pd 4$d$-Nd 4$f$ hybridization, the states near the Fermi level are found to be dominated by hybridized Nd 4$f$-Si 3$p$ states. Nd 3$d$ core level spectrum exhibits multiple features manifesting strong final state effects due to electron correlation, charge transfer and collective excitations. These results serve as one of the rare demonstrations of hybridization of Nd 4$f$ states with the conduction electrons possibly responsible for the exoticity of this compound.
\end{abstract}

\maketitle
%

%\section{INTRODUCTION}

Intermetallic compounds containing rare-earths or actinides with partially filled $f$ shells allow for the realization of complex magnetic ground states, which can be driven by the presence of competing magnetic interactions (see, for instance, see Refs. [1,2]).%\cite{Doniach,Ce}]).
The inter-site magnetic interaction among the local $f$-electrons has been known to be mainly governed by indirect exchange interaction, viz., Rudermann- Kittel-Kasuya-Yosida (RKKY) interaction in such metallic compounds. Following intense research during 1970s, it was recognized that partial delocalization of 4$f$ electrons in Ce-based materials leads to anomalous magnetism in its intermetallic alloys due to finite 4$f$-hybridization with conduction electrons (Kondo effect); exoticity in their electronic properties emerges due to the competition between the RKKY and Kondo interactions.\cite{Doniach} In fact, the 4$f$-hybridization induced effects can extend to other rare-earths, such as Pr, Nd etc. as well, was actually demonstrated through photoemission spectral features.\cite{Pr,Parks,Swapnil} This conclusion from such excited state techniques found further evidence from subsequent studies of hybridization-related ground state properties like heavy-fermion superconductivity or magnetic anomalies in some Pr and Nd based systems, the most famous ones being  skutterudites.\cite{Saha-PRB} However, the fact remains that such examples for Nd are less abundant. A notable example in the field of multiferroicity having relevance to magnetoelectric coupling is NdCrTiO$_5$, where magnetic ordering temperature is substantially enhanced \cite{Greenblatt,Hwang,Mukherjee-PRB11,Waller} with respect to isostructural Gd analogues due to 4$f$-hybridization. A recent density functional theoretical calculations, supported by inelastic $x$-scattering experiment on Nd(0001) thin film, reveal strong 4$f$ electron correlation and its influence on lattice dynamics,\cite{Waller} which stresses unexplored avenues of Nd 4$f$ hybridization. Judged by such reports, investigations of more Nd systems should present interesting situations.

In this respect, the material class, $R_2M$Si$_3$ ($R$ = rare-earth, $M$ = transition metal) is of great interest, where the $R$ sub-lattice is sandwiched by the $M$Si sublattice and the hybridization between 4$f$ and valence electrons leads to varied novel properties.\cite{Ce2CoRhSi3} These materials form in AlB$_2$-type structure; $M$ and Si atoms are placed in '$ab$' basal plane in an ordered manner making a hexagonal sublattice and form the conduction band. These planes are separated by hexagonal $R$ sublattice along '$c$'-axis.\cite{struct} In the series, $R_2$PdSi$_3$, the $R$ atoms have two different chemical environment, R1 and R2; R1 (75$\%$) has 4 Pd and 8 Si neighbors and R2 (25$\%$) is surrounded by 12 Si atoms. The space group is $P6/mmm$ or $P6_3/mnc$ depending on its supersymmetry.\cite{spacegroup} Here, Pd at the $M$ sites instead to 3$d$ transition metals has led to tremendous attraction in the recent years followed by the discovery of many exotic phenomena; such as, Ce$_2$PdSi$_3$ is a Kondo Lattice,\cite{CePd} Gd, Dy and Tb show large negative magnetoreistance,\cite{RPd} Gd$_2$PdSi$_3$ exhibits strong magneto-caloric effect (MCE) \cite{GdPd} and electrical resistivity minimum.\cite{RMallik} Quasi-one dimensional behavior is observed in the compound Tb$_2$PdSi$_3$,\cite{Paulose-TbPd} \emph{etc}. It is intriguing to note that, most recently, following the magnetic and transport investigations on Gd$_2$PdSi$_3$ [Ref. 16], %\cite{GdPd}],
even skyrmion lattice anomalies have been reported,\cite{Gd2PdSi3-Science} despite the fact that these materials form in the centrosymmetric structure.

The behavior of Nd$_2$PdSi$_3$ is very different from the other members in this series as it exhibits an unusual ferromagnetic order\cite{Rmag,sc,Ndmag} while most other members exhibit antiferromagnetic order. Moreover, the magnetic ordering occurs at a significantly higher temperature ($T_C \sim$ 16 K) than the expected value according to de Gennes scaling.\cite{Ndmag} At low temperatures, this system, in addition shows complex magnetic properties.\cite{Ndmag,neutron} The electronic properties of this compound and some of its solid solutions (substituting nonmagnetic ion at various sites) have been studied extensively using magnetization, heat capacity and resistivity measurements.\cite{Ndmag} The results indicate existence of a competing interaction between ferromagnetic and antiferromagnetic interaction at low temperatures. It was proposed that strong Nd 4$f$ hybridization with the conduction electrons plays an important role behind its unusual behavior. Here, we report our results of the study of electronic properties of Nd$_2$PdSi$_3$ using high resolution photoemission spectroscopy and electronic band structure calculation. The experimental results provide evidence for strong Nd 4$f$-hybridization.

%\section{Experimental and Calculation Details}

The single crystal of Nd$_2$PdSi$_3$ was grown and characterized as reported elsewhere.\cite{sc} Photoemission measurements were carried out using a Scienta electron analyzer, R4000 WAL and monochromatic laboratory sources of Al $K\alpha$ ($h\nu$ = 1486.6 eV), He {\scriptsize II} ($h\nu$ = 40.8 eV) and He {\scriptsize I} ($h\nu$ = 21.2 eV) photon energies. Energy resolution for ultraviolet photoemission measurements were fixed to 2 meV and for x-ray photoemission, it was 350 meV. The temperature down to 10 K was achieved by an open cycle He cryostat from Advanced Research systems, USA. The single crystal was cleaved at a base pressure of 3$\times$10$^{-11}$ torr just before the measurements. The band structure calculation was carried out using Wien2k software \cite{Wien2k} and following full potential linearized augmented plane wave method within the local density approximation (LDA). The convergence was achieved using 1000 $k$-points within the first Brillouin zone; the energy and charge convergence was achieved to be less than 0.2 meV/f.u. and 10$^{-3}$ electronic charge. The effect of spin-orbit interaction and electron-electron Coulomb repulsion was introduced in the calculation following LDA+$U$ ($U$ = electron correlation strength) method. The space group used in our calculations is $P6_3/mnc$ and lattice parameters were taken from the Ref. [22].%\cite{Ndmag}].

%\section{Results and discussion}

\begin{figure}
\vspace{-4ex}
 \begin{center}
 \includegraphics[scale=0.45]{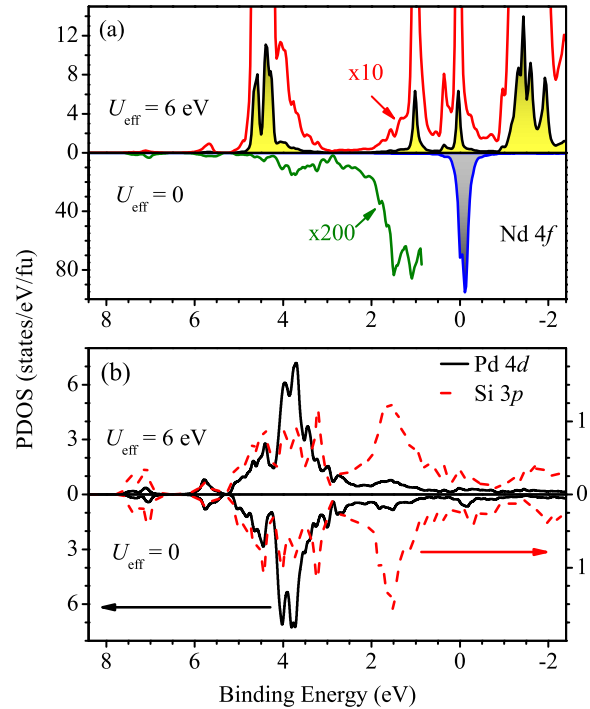}
 \end{center}
 \vspace{-18ex}
\caption{(Color online) Calculated Partial Density Of States of (a) Nd 4$f$ and (b) Pd 4$d$ and Si 3$p$ states for $U$ = 6 eV (upper panel) and $U$ = 0 eV (lower panel). Nd 4$f$ contributions are rescaled and shown in (a) to emphasize weak intensity features.}
\label{Fig1PDOS}
\end{figure}

Band structure calculation converges to a metallic ground state as also found experimentally. The valence band is found to be primarily constituted by Nd 4$f$, Pd 4$d$ and Si 3$p$ states; partial density of states (PDOS) calculated for $U$ = 0 and $U$ = 6 eV are shown in Fig. \ref{Fig1PDOS}. The calculated results exhibit significant modification for the consideration of $U$. For example, Nd 4$f$ PDOS is essentially concentrated as a narrow band at the Fermi level in $U$=0 case - weak intensities are observed at higher binding energies after we rescaled the intensities by a factor of 200. Si 3$p$ PDOS exhibit two broad bands in the energy range (0-2.5 eV) and (2.5-6 eV). On the other hand, Pd 4$d$ states primarily contribute between 2.5 - 6 eV energy regime. From the distribution of various PDOS, it is evident that Nd 4$f$ states hybridize with the Si 3$p$ states and Nd 4$f$-Pd 4$d$ hybridization appears to be weak. Pd 4$d$-Si 3$p$ hybridization leads to bonding features between 2.5-6 eV binding energies with the dominance of Pd 4$d$ contributions and the antibonding bands appear between 0-2.5 eV having large Si 3$p$ contributions.

With the inclusion of electron correlation among Nd 4$f$ electrons ($U$ = 6 eV), the scenario involving Pd 4$d$ and Si 3$p$ remain almost similar. Even, Si 3$p$-Nd 4$f$ hybridization seems to remain very similar to its $U$=0 case. But the Nd 4$f$ PDOS exhibit a strong feature near 4 eV and have substantial Pd 4$d$-Nd 4$f$ hybridization. Additionally, there are two sharp features, one at about 50 meV and the other at 1 eV.

\begin{figure}
\vspace{-4ex}
 \begin{center}
 \includegraphics[scale=0.45]{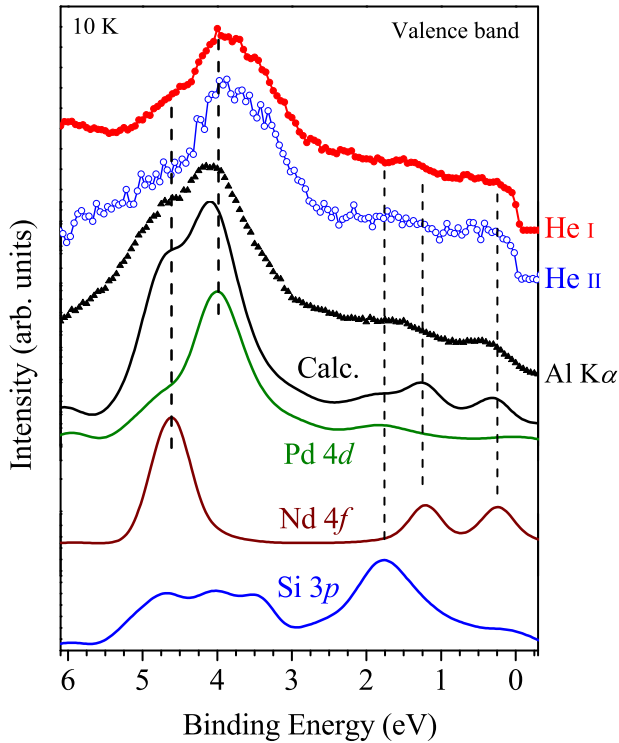}
 \end{center}
 \vspace{-18ex}
\caption{(Color online) Experimental valence band spectra (symbols) at 10 K using He {\scriptsize I} (solid circles), He {\scriptsize II} (open circles) and Al $K\alpha$ (triangles) photon sources. Calculated spectral functions and the contribution from various PDOS are also shown by lines exhibiting remarkable representation of the experimental spectra.}
\label{Fig2VB}
\end{figure}

In Fig. \ref{Fig2VB}, we show the experimental valence band spectra collected at 10 K using the monochromatic laboratory sources, He {\scriptsize I}, He {\scriptsize II} and Al $K\alpha$ after normalizing by the intensity of the most intense peak at about 4 eV binding energy. There are several features in the experimental spectra in the energy range shown down to 6 eV below the Fermi level. To identify the features, we have calculated the spectral functions from LDA+$U$ results for $U$ = 6 eV. Each of the PDOS are multiplied by the atomic photoemission cross section for Al $K\alpha$ spectroscopy.\cite{Yeh} These results are then broadened by a Gaussian representing the energy resolution and a Lorentzian representing various lifetime broadenings. The simulated total spectrum and the constituent components are shown in the figures by line exhibiting remarkable representation of the experimental Al $K\alpha$ data. It is evident that the intense feature at 4 eV binding energy is primarily contributed by Pd 4$d$ states and the feature at 4.5 eV is contributed by Nd 4$f$ states; here, rare-earth 4$f$ signal appears at binding energy higher than that of Pd 4$d$ [Ref. 26].%\cite{R2PdSi3}].

The photoexcitation cross sections of Pd 4$d$ and Si 3$p$ show similar trend of change in the photon energies used in our study. The relative photoemission cross section of Nd 4$f$ with respect to Pd 4$d$ at He {\scriptsize I} energy increases by about 3.5 times at He {\scriptsize II} energy and by about 15 times at Al $K\alpha$ energy. Therefore, the He {\scriptsize I} spectrum mainly represents the Pd 4$d$ and Si 3$p$ characters in the valence band, whereas He {\scriptsize II} and Al $K\alpha$ yield enhanced information of Nd 4$f$ state along with other contributions; Al $K\alpha$ data has the highest Nd 4$f$ contribution compared that in the other spectra. Thus, significantly smaller intensity at 4.5 eV in the UV spectra compared to the Al $K\alpha$ spectrum can be attributed to the change in relative photoemission cross section. This provides an experimental evidence of the Nd 4$f$ contribution at this energy.

The features at 0.3 eV and 1 eV are distinct in the Al $K\alpha$ data, while the intensities in the UV spectra is somewhat flat in this energy range. This indicates Nd 4$f$ contribution at these energies, which is in good agreement with the calculated results. Si 3$p$ states primarily contribute at 1.8 eV along with significant intensities in the 3-5 eV binding energy range. All these results indicate that the group of features below 3 eV binding energy are the bonding peaks and the ones between 0-3 eV are antibonding peaks arising due to hybridization between Nd 4$f$, Pd 4$d$ and Si 3$p$ states.

\begin{figure}
\vspace{-4ex}
 \begin{center}
 \includegraphics[scale=0.45]{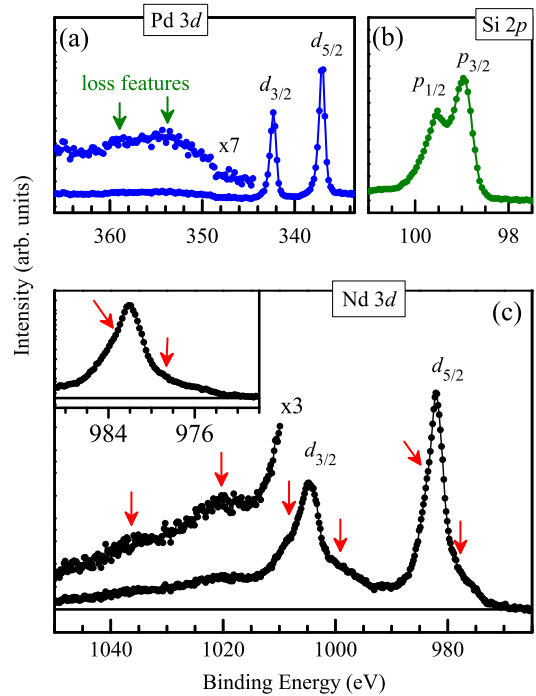}
 \end{center}
 \vspace{-18ex}
\caption{(Color online) (a) Pd 4$d$, (b) Si 3$p$ and (c) Nd 3$d$ core level spectra collected at 10 K. In (a), the extended energy region in Pd 3$d$ spectrum is rescaled in intensity to reveal loss features indicated by arrows. In (c), the arrows show the location of distinct features. High binding energy part is shown in enlarged intensity scale to emphasize the peak positions. The inset shows the Nd 3$d_{5/2}$ intense signal in an expanded energy scale.}
 \label{Fig3Core}
\end{figure}

Some selected core level spectra collected at 10 K are shown in Fig. \ref{Fig3Core}. Pd 3$d$ core level spectrum shown in Fig. \ref{Fig3Core}(a) exhibits two sharp features at about 337 eV and 342.3 eV corresponding to spin-orbit split 3$d_{5/2}$ and 3$d_{3/2}$ excitations, respectively; the intensity ratio is 3:2 as expected from the multiplicity of the features and the spin-orbit splitting is 5.3 eV. Higher binding energy regime exhibits a hump with significant intensity. For clarity, we show the spectrum with rescaled intensity exhibiting signature of two peaks (see arrows) separated by the energy of Pd 3$d$ spin-orbit splitting. This suggests their origin to be associated to plasmon-type collective excitations. The photoemission signal from Si 2$p$ core level spectra shown in Fig. \ref{Fig3Core}(b) exhibits two spin-orbit-split features at 99 eV and 99.5 eV representing 2$p_{3/2}$ and 2$p_{1/2}$ excitations, respectively; spin-orbit splitting is close to 0.5 eV. We could not probe the loss feature at the higher binding energy regime in this case due to the overlap of photoemission signals from Nd 4$d$ core level.

In Fig. \ref{Fig3Core}(c), we show the Nd 3$d$ core level spectrum, which reveals two sharp and intense peaks at 982 eV and 1004.7 eV corresponding to the spin-orbit split features 3$d_{5/2}$ and 3$d_{3/2}$ excitations, respectively. In addition, there are several features; distinct signature of the features are clearly visible in the experimental data as denoted by 'arrows' in the figure. Presence of such multiple features indicate that electron-electron interaction is finite in this system and charge transfer from the ligand levels to Nd is possible that allows screening of the photoemission core hole.

\begin{figure}
\vspace{-4ex}
 \begin{center}
 \includegraphics[scale=0.45]{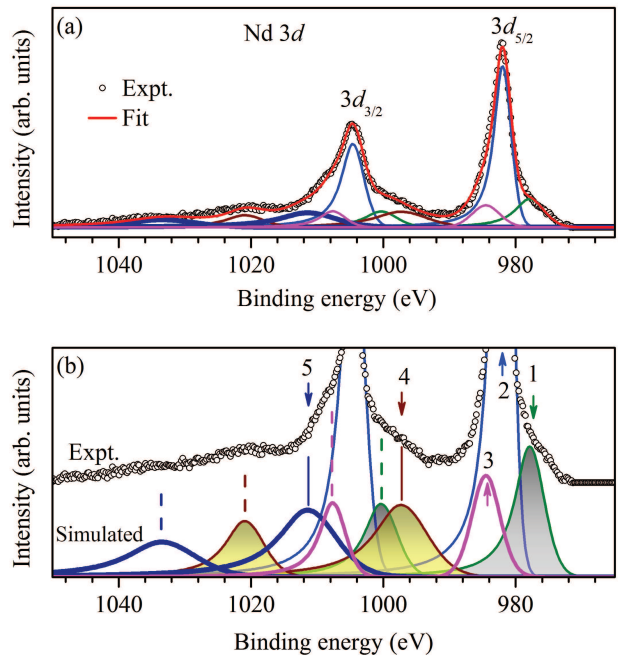}
 \end{center}
 \vspace{-18ex}
\caption{(a) Simulation of the Nd 3$d$ core level spectrum; experimental data (open circles) and simulated spectrum (red line) are superimposed over each other exhibiting remarkable representation. (b) The constituent peaks are shown in an expanded intensity scale. Arrows with the numbers represent the energy position of the features corresponding to Nd 3$d_{5/2}$ signal and the dashed vertical lines indicate features corresponding to Nd 3$d_{3/2}$ signal.}
 \label{Fig4Nd3dFit}
\end{figure}

In order to delineate the signature of these features, we have simulated Nd 3$d$ core level spectrum using a set of asymmetric functions having Doniach-{\v S}unji{\'c} lineshape.\cite{lineshape} The results are shown in Fig. \ref{Fig4Nd3dFit} exhibiting remarkable representation of the experimental spectra. Such a lineshape indicates metallic ground state that allows low energy excitations across the Fermi level and hence, asymmetry in the peak. The constituent peaks are represented by solid lines. It is evident that there are at least five features (marked by numerics in Fig. \ref{Fig4Nd3dFit}(b)) associated to the excitation of each of the spin-orbit split features.

In photoemission, the incident photon beam excites electrons from its ground state (initial state); the excited state is called the final state. The final state Hamiltonian contains all the interaction parameters present in the ground state and in addition, the core hole potential created by the excitation process. If the electrons are non-interacting, the final states will be identical to the initial states and one observes a feature corresponding to each excitations. However, if the electron-electron interaction is finite, the eigenstates of the final state Hamiltonian will be different from those for the initial state Hamiltonian and there will be finite overlap of multiple final states with the ground state wave function leading to multiple features; this is known as final state effects (see, for instance, Ref. [28]).%\cite{Nickelates}).
Assuming a configuration interaction model, the electronic configuration of the final states are as follows. Electronic configuration of Nd (Atomic number 60) is [Xe]4$f^3$(5$d^1$6$s^2$). Usually, Nd is found in (3+) valence state similar to other rare earth based materials. Therefore, the ground state will be dominated by the electronic configuration, $|4f^3>$. Due to creation of Nd 3$d$ core hole, the final states will be comprised of $|4f^3>$ states and the charge transferred states, $|4f^4\underbar{L}^1>$,  $|4f^5\underbar{L}^2>$, etc. Here, $\underbar{L}$ is a hole in the ligand band, (Pd4$d$Si3$p$).

From various core level calculations,\cite{FujimoriRMP,Nickelates,Cuprates,gunnarsson} it is well established that the lowest binding energy feature corresponds to the most stabilized final state, where the stabilization comes from the screening of the core hole with the conduction electrons. Thus, the feature `1' in the figure can be associated to the final state, $|4f^4\underbar{L}^1>$ for 3$d_{5/2}$ excitations. All other cases for 3$d_{5/2}$ excitations appear at higher binding energies; they are labelled as 2, 3, 4 and 5 with vertical arrows in Fig. \ref{Fig4Nd3dFit}. Evidently, hybridization of Nd 4$f$ with the valence electrons (mostly Si 3$p$) is finite leading to significant charge transfer induced final state features. The features `4' and `5' appear about 16 eV and 32 eV away from the main peak. In Ce-based systems, similar features are observed exhibiting features related to a Kondo resonance peak and is often attributed to Kondo physics.\cite{CeB6} However, we observe loss features in Pd 3$d$ core level spectrum shown in Fig. \ref{Fig3Core}(a) at similar energy separations, which is also consistent with the observation of plasmon excitations in other $R_2$PdSi$_3$ systems.\cite{R2PdSi3} These observations suggest that the features `4'and `5' are indeed loss features due to the plasmon excitations along with the photoexcitation of core electrons.

%\section{Conclusion}
In summary, we studied the electronic structure of Nd$_2$PdSi$_3$ using high resolution photoemission spectroscopy and ab initio band structure calculations. The comparison of the experimental valence band spectra with the calculated results provides an estimate of the electron correlation strength among Nd 4$f$ electrons close to 6 eV. The electronic states close to the Fermi level is constituted by hybridized Nd 4$f$ and Si 3$p$ states. The core level spectra exhibit multiple features indicating strong final state effects. These results provide an evidence of strong Nd 4$f$ - Si 3$p$ hybridization despite the fact that Nd 4$f$ has stronger atomic character than Ce-based systems, which could be the key for the apparent anomaly in the magnetism of this system.

%\section{Acknowledgements}
KM and EVS acknowledge financial assistance from the Department of Science and Technology, Government of India under the J.C. Bose Fellowship program and KM acknowledges the financial assistance from the Department of Atomic Energy under the DAE-SRC-OI Award.

\end{document}